\documentclass{article}
\usepackage{spconf,amsmath,graphicx,mathtools}
\usepackage{amsfonts}
\usepackage{graphicx}
\usepackage{blindtext}
\mathchardef\mhyphen="2D 
\usepackage{hyperref}
\usepackage{color}
\usepackage{cite}

\title{BEAST: ONLINE JOINT  BEAT AND DOWNBEAT TRACKING BASED ON STREAMING TRANSFORMER}
\name{Chih-Cheng Chang and Li Su}
\address{Institute of Information Science, Academia Sinica, Taipei, Taiwan}
\begin{document}
%
\maketitle
\begin{abstract}
Many deep learning models have achieved dominant performance on the offline beat tracking task. However, online beat tracking, in which only the past and present input features are available, still remains challenging. In this paper, we propose \textbf{BEA}t tracking  \textbf{S}treaming \textbf{T}ransformer (BEAST), an online joint beat and downbeat tracking system based on the streaming Transformer. To deal with online scenarios, BEAST applies contextual block processing in the Transformer encoder. Moreover, we adopt relative positional encoding in the attention layer of the streaming Transformer encoder to capture relative timing position which is critically important information in music. Carrying out beat and downbeat experiments on benchmark datasets for a low latency scenario with maximum latency under 50 ms, BEAST achieves an F1-measure of 80.04\% in beat and 46.78\% in downbeat, which is a substantial improvement of about 5 percentage points over the state-of-the-art online beat tracking model.
\end{abstract}
\section{Introduction}
\label{sec:intro}
Music beat tracking is a fundamental task 
in the field of music information retrieval (MIR), as beats are the primary incentive for people  
to get involved and interact with music. Music beat tracking is also a task that is related to the musical semantics in both the 
temporally local level (e.g., individual sound events, articulations) 
and the global level (e.g., tempo, meter). 
In recent years, the advancement of deep learning techniques has achieved dominant performance on the conventional offline beat tracking task 
\cite{beat3,beatcrnn,MCTbeat}. Variants of models, such as 
recurrent neural networks (RNNs) \cite{BockFF}, temporal convolutional networks (TCNs) \cite{bock2020}, as well as Transformers \cite{BeatTrans,beattrans2}, have all been found effective in offline beat tracking, due to the capability of these models in 
exploiting both global context and local sound events information simultaneously.

Among these models, the Transformer is undoubtedly the one which receives the most attention \cite{Attention}.
Transformer, or more specifically, the multi-head attention (MHA) mechanism, is a sequence modeling technique which has achieved state-of-the-art performances in numerous tasks ranging from the 
natural language processing (NLP) to MIR fields, 
as it leverages a combination of information from different positions of the input, 
a way suitable for exploiting both local and global information. However, similar to other sequence modeling techniques such as the bidirectional RNN, the Transformer 
does not support online processing: the entire input sequence is required to compute the attention. This hinders the use of Transformers in many real-time or interactive MIR applications \cite{aubio,interact}, in which online beat tracking represents a technical barrier.

Online beat tracking is less discussed than offline beat tracking. 
An early work on this task 
performed machine learning to 
extract onset strength signals and locate beat positions based on weighted autocorrelation \cite{aubio}. 
Another approach involves inferring beat positions in real-time using multi-agent models \cite{IBT}, which initialize a set of agents with various hypotheses that try to validate their respective hypotheses based on observations along time. However, these knowledge-based approaches are prone to noise in audio signals and are not able to achieve promising performance. With the growing success of deep learning, data-based beat tracking methods have become more prominent. Some recent works employed causal models such as unidirectional RNNs and convolutional RNNs (CRNNs) to obtain beat activations for online beat tracking \cite{Beatnet,1Dstate}. Although the deep learning models have outperformed the conventional knowledge-based model, the performance is still much worse compared to the offline beat tracking systems.

In this paper, we propose BEAST\footnote{\href{https://github.com/WildHoneyPie/BEAST/}{https://github.com/WildHoneyPie/BEAST/}}, a novel online joint beat and downbeat tracking based on streaming Transformer encoder \cite{CBP}. Based on the contextual block processing mechanism in the encoder, the output can be produced by giving partially available music audio frames rather than the entire input sequence. To further outperform, rather than the absolute positional encoding which is used in the original contextual block processing encoder, inspired by transformer-XL  \cite{TransformerXL} and music transformer \cite{Musictransformer}, BEAST adopts the relative attention mechanism in the streaming transformer encoder.

To verify the performance of the proposed method, we carry out experiments on online beat and downbeat tracking. For a low latency scenario with a 46 ms latency, BEAST substantially outperforms all the existing online beat and downbeat tracking systems. According to our knowledge, this is the first work to utilize the streaming transformer in MIR.

\section{Related work}
\label{sec:relation}
\subsection{Beat tracking}
\label{ssec:beattracking}
In general, beat tracking approaches can be structured as a three-stage sequential learning task. The first stage extracts the raw audio signal to features such as spectrogram or chromagram that represent the content of the musical signal. The second step consists of a feature learning stage that aims to determine the likelihood of beat activation among the features extracted in the previous step. Finally, a post-processing stage consisting of a probabilistic graphical model is used to pick up the beat sequence from the raw model output.

There have been several deep learning approaches to determine the likelihood of beat activations for offline beat tracking. 
Besides, multi-task learning further makes a breakthrough on beat tracking. Beat, downbeat, and tempo are strongly correlated features. Sharing model weights among all the three tasks helps each to reach better convergence \cite{bock2020}. However, offline beat tracking systems use non-causal DNN models that rely on the entire input sequence, which is not suitable for online beat tracking. 

Some methods have been proposed by adopting the uni-directional RNN which operates on a stream of input features for online beat tracking \cite{BockFF}. At the post-processing stage, probabilistic models such as the dynamic Bayesian network (DBN) or particle filter that do not require complete beat sequence are used to determine the most likely beat location \cite{Beatnet,1Dstate}.
In this work, we inherit the fashion of multi-task learning and DBN, while substituting the uni-directional RNN with a novel streaming Transformer encoder.

\subsection{Streaming Transformer}
\label{ssec:streaming Trans}
Transformer has a drawback in that the entire input sequence is required to compute self-attention, making it difficult to utilize in online processing systems.
Dong et al. \cite{Selfalign} introduced a chunk-hopping mechanism to the CTC-Transformer model to support online processing. The block processing simply solves the problems and is suitable for online processing. However, it loses the global context information and degrades performance in general.

To prevent performance degradation due to block processing, Tsunoo et al.  \cite{CBP} utilize the additional context embedding that is handed over from the preceding block to the following one. Augmented memory transformer (AM-TRF) uses the block processing method with an additional memory bank to reduce the computation in capturing the long-range left context\cite{AMTRF}.
In this work we inherit the block processing mechanism in \cite{CBP} .
\begin{figure}[t]
\centering
\centerline{\includegraphics[width=9cm]{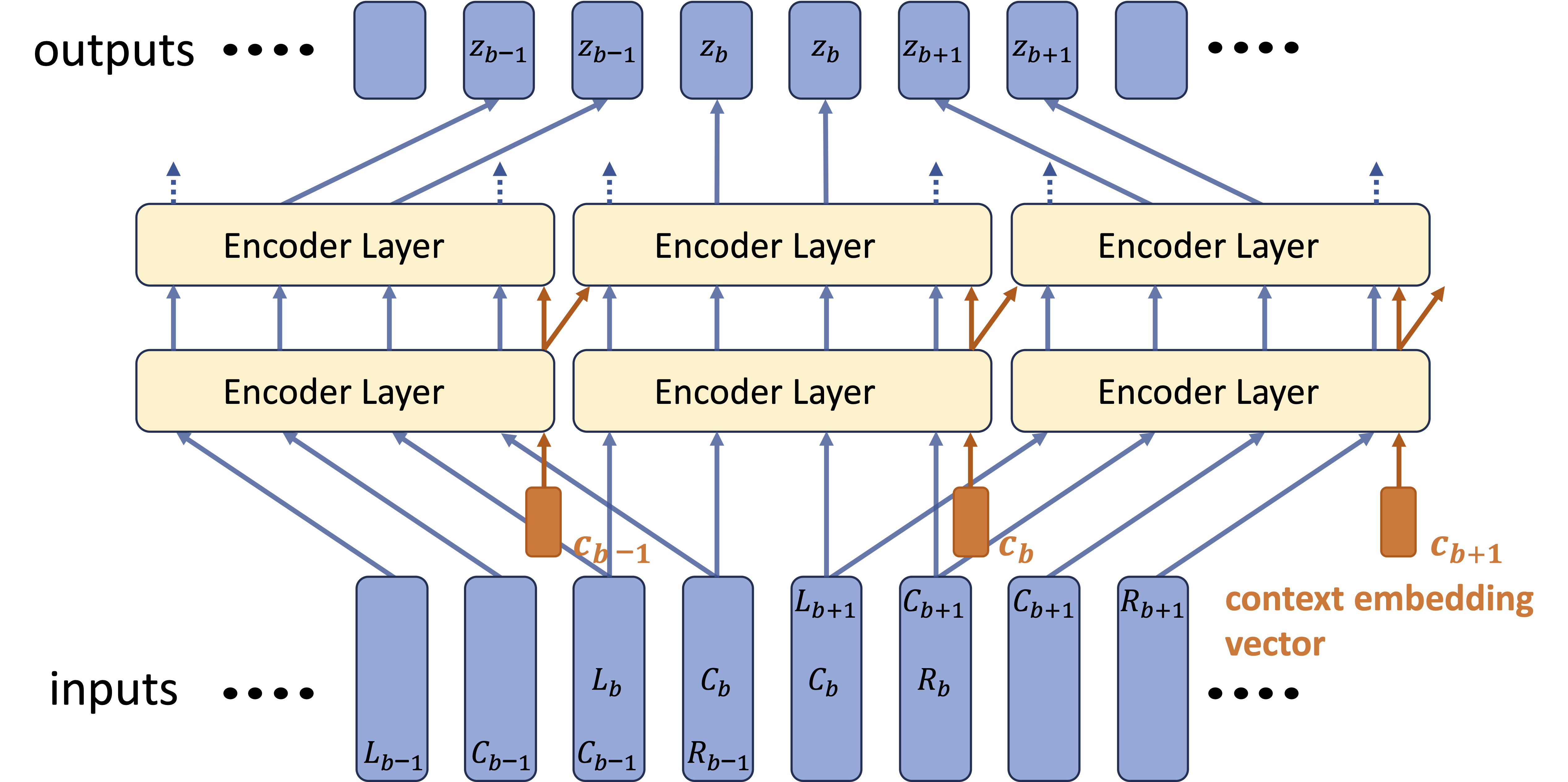}}

\caption{The contextual block processing encoder.}
\label{figure1}

\end{figure}

\section{Methodology}
\label{sec:Method}

\subsection{Contextual block processing Transformer}
\label{ssec:CBP}
For the vanilla Transformer encoder, 
we denote the output feature of its $n^{th}$ layer as $z^n \in \mathbb{R}^{T\times d}$, where $T$ and $d$ are the input sequence length and feature dimension, respectively. Denote the projection matrices of query ($Q$), key ($K$), and value ($V$) as $W^n_Q$, $W^n_K$, and $W^n_V$, respectively. The projection operation at the $n^{th}$ layer is then represented as $Q^n := z^{n-1} W^n_Q$, $K^n := z^{n-1} W^n_K$, and $V^n := z^{n-1} W^n_V$. The attention scores $A^n$ computed by the dot product of queries and keys are then represented as 
\begin{align}
A^n := Q^nK^{n\top} = z^{n-1} W^n_Q {W^n_K}^\top {z^{n-1}}^\top\,,
\end{align}
and the attention weights $a^n$ are represented as 
\begin{align}
a^n := &\,\,\text{Attention}(Q^n,K^n,V^n) \nonumber\\
    :=& \,\, \text{softmax}\left(\frac{Q^nK^{n\top}}{\sqrt{d}}\right)V^n\,.
\end{align}

\begin{figure*}[htb]
\centering
\centerline{\includegraphics[width=14.1cm]{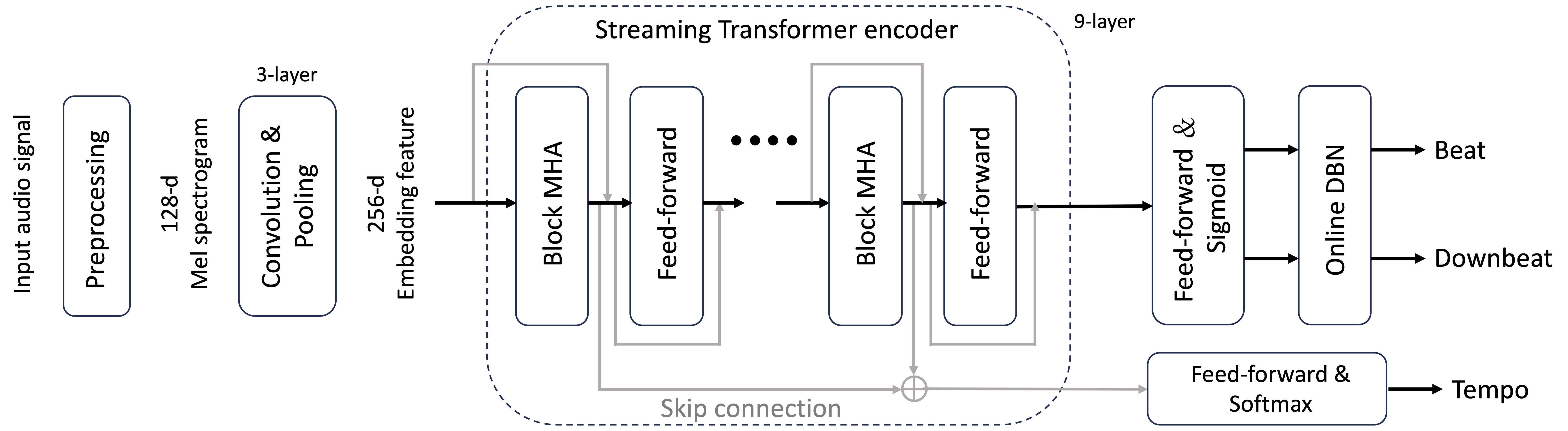}}
\caption{The complete model architecture of BEAST.}
\label{figure2}
\end{figure*}

Different from the vanilla Transformer which operates directly on the whole sequence, the streaming Transformer utilizes block processing with a context inheritance mechanism \cite{CBP}. First, the sequence of input feature vectors is chunked into multiple non-overlapping blocks $C:=\{C_1, C_2, \cdots, C_b, \cdots\}$, where $b$ is the block index
and each block contains $N_c$ frames.
In order to reduce the boundary effect in forming the contextual block, for each $C_b$ we introduce the left and right sub-blocks, namely $L_b$ and $R_b$, respectively, where $L_b$ is an $N_l$-frame blocks before $C_b$, and $R_b$ is an $N_r$-frame block after $C_b$. The number $N_l$, $N_c$ and $N_r$ are all positive integers. Concatenating them 
forms a contextual block $z^n_b = [L_b, C_b, R_b]$ for the input of the Transformer encoder. Fig. 1 illustrates a simple example where $N_c=2$ and $N_l=N_r=1$.
As shown in Fig. 1, 
when a block is fed into the encoder, the features of interest 
are processed for the output with right contexts provided by the look-ahead frames, as well as left contexts provided by history frames. Besides, for each block $z^n_b$, an additional context embedding vector $c^n_b$ inherited from the previous block is utilized in each layer of each block to capture the long-term dependency of the input sequence. The details about the computation of $c^n_b$ can be found in the related work \cite{CBP}. 

With the additional context embedding vector of block $n$ and layer $b$ as $c^n_b$, the input queries, keys, and values in self-attention layer are reformulated as $q^n_b = [z^{n-1}_b , c^{n-1}_b]$ and $k^n_b = v^n_b = [z^{n-1}_b , c^{n-1}_{b-1}]$, and the computation of the attention weights in (2) of each block and layer are rewritten as
\begin{gather}
a^n_b := \text{Attention}(Q^n_b ,\ K^n_b,\ V^n_b)\,, \label{eq:attention}
\end{gather}
where $(Q^n_b,\ K^n_b,\ V^n_b) = (q^n_bW^n_Q,\ k^n_bW^n_K,\ v^n_bW^n_V)$.
We then apply the contextual block processing Transformer encoder to extract the beat activations for online beat tracking.

\subsection{Relative positional encoding}
\label{ssec:RPE}
Both the vanilla Transformer and original contextual block processing streaming transformer apply positional sinusoids to the input embedding to encode absolute position information \cite{Attention,CBP}. However, music has multiple dimensions along which relative differences matter more than their absolute values. To capture pairwise positional relationships between input elements, several methods have modulated the self-attention with a position relation-aware mechanism \cite{Musictransformer,TransformerXL}.

We apply the relative positional encoding introduced in \cite{TransformerXL} in our MHA layer. With additional bias terms for queries and the sinusoid formulation for relative position encoding, the attention scores of two input sequence arbitrary positions $i$ and $j$ in each block are 
\begin{gather}
\begin{split}
A^n_{b,i,j} := &Q^n_{b,i} {K^n_{b,j}}^\top + Q^n_{b,i}  {W^n_R}^\top {R_{i-j}}^\top \\
&+ u{K^n_{b,j}}^\top + v{W^n_R}^\top {R_{i-j}}^\top\,, \label{eq:relative_position_encoding}
\end{split}
\end{gather}
where the newly added item $R_{i-j}$ is the sinusoid encoding vectors that provide the prior of relative position, and $W^n_R \in \mathbb{R}^{d\times d}$ is a trainable matrix projecting $R_{i-j}$ into a location-based key vector. The two bias terms $u, v \in \mathbb{R}^{d}$ are both trainable vectors. Applying the attention scores in (\ref{eq:relative_position_encoding}), the attention weights in (\ref{eq:attention}) 
are rewritten into 
\begin{gather}
a^n_b :=  \text{softmax}\left(\frac{A^n_b}{\sqrt{d}}\right)V^n_b\,.
\end{gather}

\subsection{Complete architecture}
\label{ssec:CA}
Fig. 2 shows the complete architecture of BEAST. 
Each audio file (with a sampling rate of 44.1kHz) is first transformed into a log-magnitude spectrogram with a hop size of 23 ms (1024 samples) and a window size of 93 ms (4096 samples). Logarithmic grouping of frequency provides an input representation with a total of 128 frequency bands from 30 Hz up to 11 kHz. The three 2D convolutional and max pooling layers as the front-end feature extractor map the 128-dimensional input features to 256-dimensional vectors.

BEAST comprises 9 contextual block processing streaming Transformer encoder layers. Each encoder layer has eight heads of block processing self-attention network followed by a position-wise feed-forward network with hidden dimension = 1024, both of which have residual connections. Layer normalization is performed before each network. Following multi-task learning practice, we utilize a feed-forward layer to map the representation to beat and downbeat activations respectively, and add a regularization branch predicting global tempo via skip connections \cite{bock2020}. We apply DBN in the madmom package \cite{madmom} at the post-processing stage. Since we do not have the complete beat sequence for the online scenario, we cannot use Viterbi decoding to obtain the global best solution. Instead, we use the forward algorithm to determine the beat sequence \cite{ROBOD}. We optimized for the DBN parameters $observation\_lambda$ and $transition\_lambda$ that achieve the best performance for each model.

\section{Experiments}
\label{sec:experiment}

\subsection{Datasets and setup}
\label{ssec:data}
We use a total of 5 datasets:  \emph{Beatles} \cite{beatles}, \emph{Ballroom} \cite{ballroom}, \emph{Hainsworth} \cite{hains}, \emph{Carnetic} \cite{carnetic}, and \emph{SMC} \cite{smc} for training and evaluated in an 8-fold cross validation manner, and \emph{GTZAN} \cite{gtzan} is used for testing only.

Beat and downbeat annotations are each represented as a 1D binary sequence that indicates beat (1) and non-beat (0) states at each input frame. The joint tracking with the tempo branch generally benefits the performance on beat tracking, as it may serve as a regularization term that helps reach better convergence.

For training, we equally weigh the binary cross entropy loss over beat, downbeat, and tempo to make a combination of them. We use a batch size of 1 to train whole sequences with different lengths. For excessively long audios, we split them into 30-second (1.3k-frame) clips. We apply RAdam and Lookahead optimizer with an initial learning rate of 1E-3, which is reduced by a factor of 5 whenever the validation loss gets stuck for 6 epochs before being capped at a minimum value of 1E-7. We use dropout with a rate of 0.5 for the tempo branch and 0.1 for other parts of the network. Our model has 7.5M trainable parameters and is trained using an NVIDIA GeForce RTX 4090 GPU with fp32 precision. We trained all the models for 50 epochs to reach sufficient convergence and picked the checkpoint with the lowest validation loss as the best model. 

We compute different metrics to evaluate BEAST, capturing two important aspects of online systems: the performance and the latency. We use the F1-measure with a tolerance window of ±70 ms to evaluate online beat and downbeat tracking performance on the GTZAN dataset. In terms of latency, we focus on the algorithmic latency induced by the streaming Transformer encoder, which equals to the center block frame size plus the right look-ahead frame size. To measure real-time factors (RTFs), we use hosts with Intel i9-12900K 16-core CPU with 3.2 GHz clock. In measuring RTFs, the beat position of a 30-second audio is concurrently estimated.

\subsection{Results}
\label{ssec:eval}
We first compare three ablation models to verify the effect of positional encoding in the Transformer for beat tracking. The first model has no positional encoding in the self-attention encoder layer. The second model utilizes absolute positional encoding into input embedding. The last model utilizes relative positional encoding with a position relation-aware self-attention mechanism. The left, center, and right context frame size ($N_l, N_c ,N_r$) of the segmented block in the encoder is set as (256, 16, 16). In Table 1, we can observe the model with relative positional encoding yields substantially better performance than the other two models.

\begin{table}[t]
\begin{tabular}{|lc|}
\hline
Model                           & F1-measure beat (\%) \\ \hline
w/o positional encoding         & 79.48                                                          \\
w/ absolute positional encoding & 80.52                                                          \\
w/ relative positional encoding & \textbf{83.65}                                                          \\ \hline
\end{tabular}
\caption{\label{table1} Effect of positional encoding on F1-measure of online beat tracking.}
\end{table}

One key issue for the online beat tracking is to compromise between latency and performance. In the following experiments, we keep the left context size $N_l$ = 256. To investigate the effect of segment length and context size, we compare the performance with different algorithmic latency by varying the $N_c$ and $N_r$. Based on our experience, keeping the $N_c$ and $N_r$ the same can achieve better performance with the latency constraints. Table 2 shows the comparison of BEAST models with different latency. We name the model with different look-ahead sizes $N_r$ as ``BEAST-$N_r$". The latency of existing models is defined as the hop size of the short-time Fourier transform (STFT) for feature extraction.

For a medium latency of 743 ms, BEAST obtained over relative 9\% and 13\% F1-measure improvement over the state-of-the-art online beat and downbeat tracking models based on CRNN \cite{1Dstate}, respectively. For a low latency of 46 ms, BEAST still obtained relative 5\% improvement on beat tracking. Because medium latency beat tracking chunks audio into fewer segments, low latency beat tracking gives higher RTF than medium latency beat tracking. While the RTFs of BEAST are higher than other existing online beat tracking models due to the more computational resources on Transformer, the RTFs are still much less than 1 and suitable for online scenarios.
\begin{table}[t]
\begin{tabular}{|l@{\hspace{0.1cm}}c@{\hspace{0.15cm}}c@{\hspace{0.15cm}}c@{\hspace{0.15cm}}c|}
\hline
Model       & \begin{tabular}[c]{@{}c@{}}F1-measure \\ beat (\%)\end{tabular} & \begin{tabular}[c]{@{}c@{}}F1-measure \\ downbeat (\%)\end{tabular} & \begin{tabular}[c]{@{}c@{}}Latency \\ (ms)\end{tabular} & RTF  \\ \hline
\multicolumn{5}{|c|}{Online Methods}                                                                                                                                                                                 \\ \hline
BEAST-$1$     & 80.04                                                           & 46.78                                                               & 46                                                      & 0.41 \\
BEAST-$2$     & 81.27                                                           & 47.23                                                               & 93                                                      & 0.21 \\
BEAST-$4$     & 82.88                                                           & 51.42                                                               & 186                                                     & 0.11 \\
BEAST-$16$    & \textbf{83.65}                                                         & \textbf{52.54}                                                               & 743                                                     & 0.08 \\
IBT \cite{IBT}         & 68.99                                                           & —                                                                   & 23                                                      & 0.16 \\
Böck FF \cite{BockFF}    & 74.18                                                           & —                                                                   & 46                                                      & 0.05 \\
BeatNet \cite{Beatnet}     & 75.44                                                           & 46.49                                                               & 20                                                      & 0.05 \\
Novel 1D \cite{1Dstate}   & 76.48                                                           & 42.57                                                               & 20                                                      & 0.02 \\ \hline
\multicolumn{5}{|c|}{Offline Methods}                                                                                                                                                                                \\ \hline
Transformer \cite{BeatTrans} & 88.5                                                            & 71.4                                                                & —                                                       & —    \\
TCN \cite{bock2020}         & 88.5                                                            & 67.2                                                                & —                                                       & —    \\ \hline
\end{tabular}
\caption{\label{table2} Performance and latency comparison of BEAST with other beat and downbeat tracking models.}

\end{table}

\section{Conclusion}
\label{sec:Conclusion}

In this paper, we have introduced BEAST, a novel online joint beat and downbeat tracking system based on streaming Transformer with the contextual block processing mechanism. We have demonstrated the effectiveness of BEAST in online beat and downbeat tracking according to its substantial improvement over the state-of-the-art models. Thanks to the block processing mechanism, the latency of BEAST is also controllable and can be adapted to different application scenarios.
This is the first demonstration of streaming Transformer in MIR. The streaming Transformer architecture can also be applied to other MIR tasks, 
such as real-time transcription or the generative model for real-time accompaniment systems. These will be left as our future work.


\vfill
\pagebreak

\clearpage
\newpage
\nocite{*}
\bibliographystyle{IEEEbib.bst}
\bibliography{refs.bib}


\end{document}